\documentclass{article}

\usepackage[final]{neurips_2019}

\usepackage[utf8]{inputenc} 
\usepackage[T1]{fontenc}    
\usepackage{hyperref}       
\usepackage{url}            
\usepackage{booktabs}       
\usepackage{amsfonts}       
\usepackage{nicefrac}       
\usepackage{microtype}      
\usepackage{amsmath,amssymb}
\usepackage{subcaption}
\usepackage{soul}
\usepackage{makecell}
\usepackage{graphicx} 

\title{Synthetic Epileptic Brain Activities using GANs}

\author{%
  Dami\'an Pascual \\
  ETHZ \\
  dpascual@ethz.ch
   \And
  Amir Aminifar\\
   EPFL \\
  \AND
  David Atienza \\
  EPFL \\
   \And
   Philippe Ryvlin\\
   CHUV \\
   \And
   Roger Wattenhofer \\
   ETHZ \\
}

\begin{document}

\maketitle

\begin{abstract}
  Epilepsy is a chronic neurological disorder affecting more than 65 million people worldwide and manifested by recurrent unprovoked seizures. Modern systems monitoring electroencephalography (EEG) signals are being currently developed with the view to detect epileptic seizures in order to alert caregivers and reduce the impact of seizures on patients' quality of life. Such systems use machine learning algorithms that require large amounts of labeled seizure data for training. However, acquiring EEG signals of seizures is a costly and time-consuming process for medical experts and patients. In this work, we generate synthetic seizure-like EEG signals, that can be used to train seizure detection algorithms, alleviating the need for recorded data. First, we train a Generative Adversarial Network (GAN) with data from 30 epilepsy patients. Then, we generate synthetic training sets for new, unseen patients, which overall yield comparable detection performance to real-data training sets. We demonstrate our results using the datasets from the EPILEPSIAE Project, one of the world's largest public databases for seizure detection. 
\end{abstract}

\section{Introduction}
Epilepsy is the fourth most common chronic neurological disorder worldwide \citep{hirtz2007common}, affecting over 65 million people. Epilepsy manifests itself by recurrent unprovoked seizures due to abnormal activity in the brain. The length of the seizures can range from few seconds to several minutes with a large variety of symptoms, including sensory auras, loss of awareness, automatic movements and full body convulsions \citep{blumenfeld2012impaired}. These symptoms not only degrade the quality of life of the patients, but they are also associated with a mortality rate 5 times higher among patients with recurrent seizures \citep{sperling1999seizure} than in the corresponding group of the general population. One third of epilepsy patients suffer from drug-resistant uncontrolled seizures, which time of occurrence is usually unpredictable. A promising solution to reduce mortality and to improve the living standard of epilepsy patients is continuous real-time monitoring using wearable technologies that collect and process EEG signals from the patient and, upon occurrence of seizures, raise alerts to caregivers or family members\citep{goverdovsky2017hearables,sopic2018glass,debener2015unobtrusive}. 

However, a barrier in developing reliable epileptic seizure detection systems is a lack of sufficient training data. Indeed, modern detection systems are driven by machine-learning-based algorithms \citep{alotaiby2014eeg} that require a considerable amount of recorded seizures in order to reliably detect future seizures. Collecting and labeling EEG data from epilepsy patients is a costly process that currently requires in-hospital recording in specialized units. Such recordings are performed in clinical practice in a minority of patients and over short periods of time, typically a week, enabling to only record a few seizures per patient. This is a major limitation considering the privacy concerns that exist around sharing medical data and the current trend towards personalized medicine. As a consequence, it is necessary to acquire significant amounts of new data for each patient. 

The problem of scarce reliable training data is common in the field of artificial intelligence and it is particularly severe in the specific case of epilepsy monitoring. The most comprehensive solution to this problem consists of generating synthetic data that can be used to train the detection algorithms. However, generating high quality medical data is challenging, and only recently substantial progress has been made thanks to advances in deep generative models and in particular, Generative Adversarial Networks (GANs) \citep{goodfellow2014}. Several studies have used synthetic data in areas such as medical imaging \citep{frid2018gan,shin2018medical,costa2018end} and Intensive Care Unit (ICU) monitoring \citep{esteban2017real,che2017boosting, lipton2016modeling} to augment existing training sets in order to improve detection accuracy. Although this data augmentation approach has proved effective, previous attempts to train only with synthetic data have reported such a strong degradation in performance \citep{esteban2017real, shin2018medical} that it has not been possible so far to dispense with real training data. Therefore, the scenario where no real training data can be accessed and only a purely synthetic training set may be available remains unsolved. This is, however, a common scenario in several medical applications, including epilepsy, given the difficulties and privacy concerns associated with collecting and sharing medical data \citep{price2019privacy}. 

The generation of synthetic EEG data has not been extensively studied in the existing literature and previous work on synthetic EEG generation has not tackled the generation of epileptic seizures \citep{hartmann2018eeg, aznan2019simulating, corley2018deep}. In this work, we present the use of a GAN to produce high quality synthetic epileptic seizure electroencephalogram (EEG) signals that can be used to train detection algorithms and achieve state-of-the-art results. 


\section{Generative model}

Our model is a conditional GAN\citep{mirza2014conditional} that, given non-seizure (inter-ictal) EEG samples at the input, generates EEG samples of epileptic seizures (ictal). The rationale behind our design is that, while epileptic seizures are very costly to record, inter-ictal signals can be easily recorded. As a result, we condition the network on inter-ictal samples from the target patient in order to provide additional information to the generator that can be exploited to produce more realistic seizure samples. In this way, we can use an already existing database to train our GAN and then use the GAN to generate seizure samples for a new patient. 

The generator of our model is a U-net \citep{ronneberger2015u} convolutional autoencoder network with weighted skip connections where the decoder translates the latent code into an ictal sample. In order to introduce stochasticity into the model, gaussian noise with mean 0 and standard deviation 1 is concatenated to the latent code. The skip connections multiply the feature maps at each layer of the encoder with a weight which is learnt during training, and then, the result of that operation is added to the corresponding feature map of the decoder. The discriminator of our GAN has the same structure as the encoder of the generator, but it includes an additional fully connected layer at the output. The loss function of our GAN is based on the Least Squares GAN (LSGAN) \citep{Mao_2017_ICCV}. Consequently, the minimization objective of the discriminator is given by:
\begin{equation}
\begin{aligned}
& \underset{\theta_D}{\text{min}}
& & \mathcal{L}_D(\theta_D) = \mathbb{E}_{\mathbf{x}\sim{p}}[(D(\mathbf{x};\theta_D) - 1)^2] + \mathbb{E}_{\mathbf{x}\sim{p}}[(D(G(\mathbf{x});\theta_D))^2],\\
\end{aligned}
\end{equation}
where the function $D$ corresponds to the discriminator and $G$ to the generator and. The input data $\mathbf{x}$ is sampled from the input data distribution $p$, and $\theta_D$ are the network parameters of the discriminator. On the other hand, the loss function of the generator, with network parameters $\theta_G$, is given by:

\begin{equation}\label{Gen_loss}
\begin{split}
\min_{\theta_G} \mathcal{L}_G(\theta_G) = \mathbb{E}_{\mathbf{x}\sim{p}}[(D(G(\mathbf{x};\theta_G)) -1)^2] +
\lambda ||G(\mathbf{x};\theta_G) - \mathbf{y}||_1,\\
\end{split}
\end{equation}

The generator's loss includes a weighted $L_1$ regularization term that ensures that the generated signal is similar to the reference output signal $\mathbf{y}$, which makes the training more stable. In Equation \ref{Gen_loss}, $\lambda$ is a hyperparameter that we fix to 100 in order to scale both terms of the loss function to a comparable magnitude, preventing the regularization term from dominating the optimization problem. The first term of the generator's loss encourages the generator to produce synthetic samples that are classified as 1, i.e., real, by the discriminator, which is adversarial with respect to the discriminator's loss function. Hence, the competing interests of the generator and the discriminator during training drive the generator to produce more and more realistic samples. 

\section{GAN training}
To train our model, we used data from the EPILEPSIAE project database \citep{ihle2012epilepsiae}, which is one of the world's largest public databases for seizure detection. The dataset contains recordings from 30 different epilepsy patients with a total of 277 epileptic seizures that sum to a duration of 21,001 seconds altogether. The EEG data is collected at a sample frequency of 256 Hz and it is divided into recordings of one hour, each one corresponding to one recording session. 

In this work, we target the setup of real-world and stigma-free wearable monitoring devices \citep{hoppe2015novel} and thus we consider only the electrodes F7T3 and F8T4 in the standard 10--20 system\citep{klem1999ten}, which can be easily hidden in glasses. We extract samples of four seconds of duration, since this length is effective to detect epileptic seizures. Given that the data was recorded at a frequency of 256 Hz, this results in samples of length 2048, i.e., 1024 per electrode. To construct the training set, we pair each ictal sample to an inter-ictal sample from the same patient. In this manner, the generator learns to map inter-ictal samples to ictal samples for any given patient. 
In order to train the GAN, the leave-one-out strategy is followed: for each target patient, the GAN is trained using the ictal and inter-ictal data coming from all other patients. The exact number of training samples depends on the number of seconds of seizure recording available in the database for all patients except for the left-out patient and, although it varies slightly, it is approximately 20,000 samples. Following this scheme, the GAN is trained independently for each patient and thus, we obtain one model per patient.

\section{Evaluation of Synthetic Data}
\begin{figure}
\begin{subfigure}{.5\textwidth}
  \centering
  \includegraphics[width=0.8\linewidth]{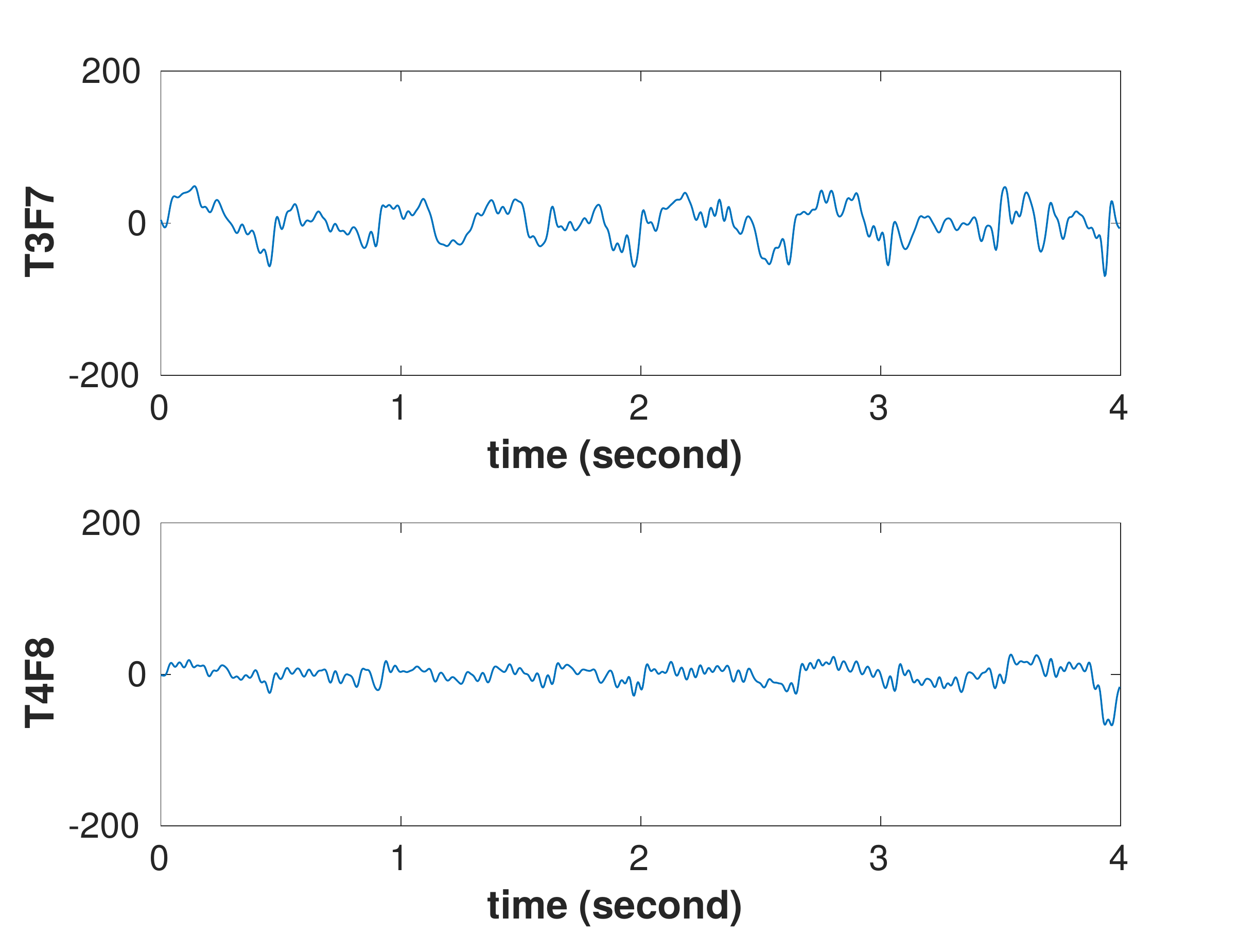}
  \caption{4 seconds of a real EEG ictal sample}
  \label{fig:sfig1}
\end{subfigure}%
\begin{subfigure}{.5\textwidth}
  \centering
  \includegraphics[width=0.8\linewidth]{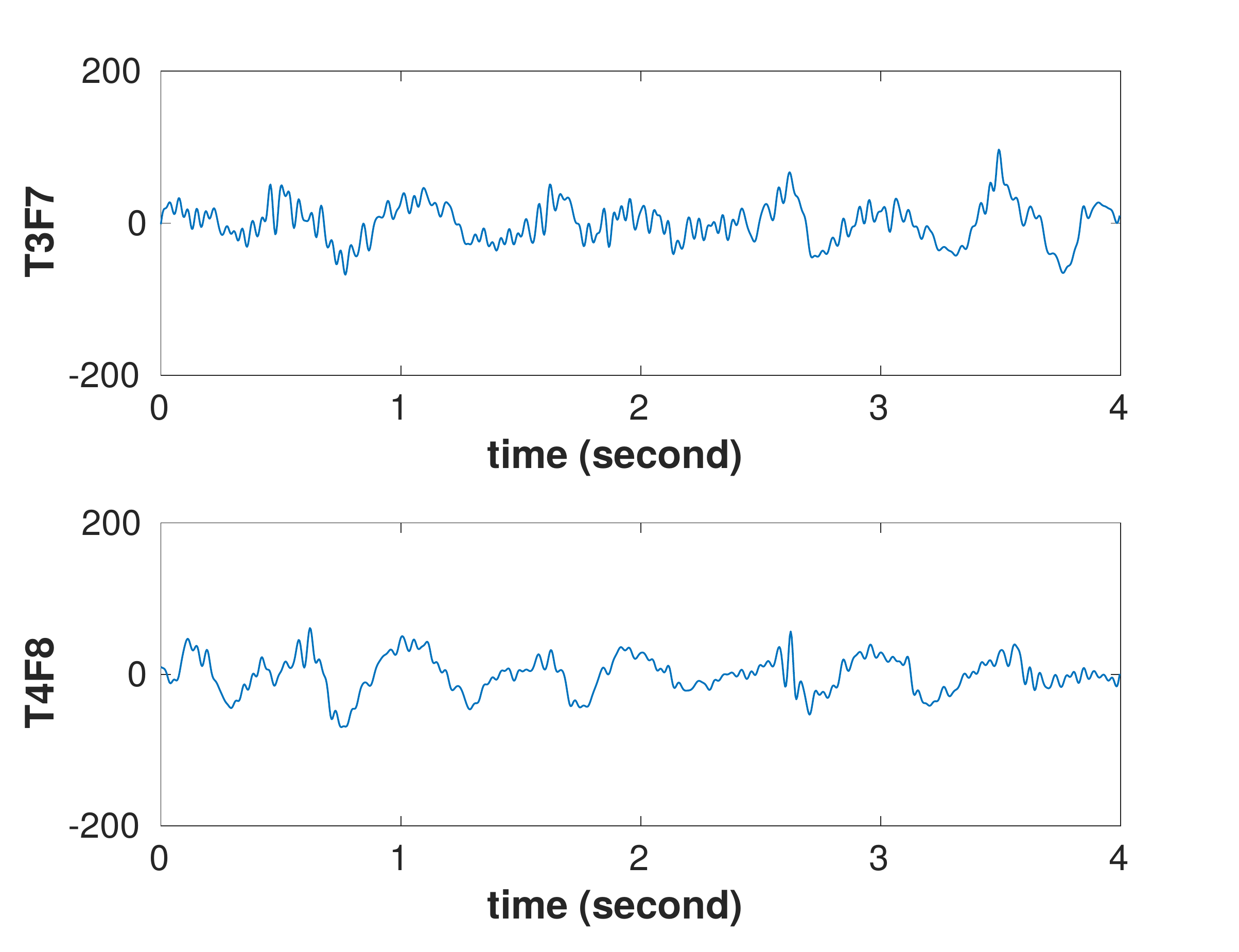}
  \caption{4 seconds of a synthetic EEG ictal sample}
  \label{fig:sfig2}
\end{subfigure}
\caption{Comparison of a real and a synthetic ictal sample for electrodes T3F7 and T4F8. It is visible how in both cases the delta-theta rhythm are present in the signals.}
\label{seiz}
\end{figure}

For each trained model, we generate between 2,000 and 6,000 ictal samples from inter-ictal EEG signals from the patient that was left out during training. In Figure~\ref{seiz}, some of the generated samples are shown in the time domain. The presence of the well-known delta--theta rhythm, i.e., rhythmic slow activity with a frequency of oscillation in 0.5--4 or 4--7 hertz, is a clear indication of the correct generation of the ictal discharge and epileptic seizure segment in the synthetically-generated EEG signals~\citep{osorio2016epilepsy}.

In order to evaluate the quality of the generated ictal samples beyond their visual appearance, we use them to train a state-of-the-art classifier based on the random forest algorithm~\citep{diaz2006gene}. The task of the classifier is to determine whether an incoming four-second sample is an ictal or an inter-ictal sample. Here, we follow the experiments performed in \cite{sopic2018glass}, which are tailored to a stigma-free wearable device for epilepsy monitoring.

We target each patient independently and consider the scenario where the only data available for training the seizure detection algorithm are real inter-ictal samples and synthetic ictal samples from the target patient. As a baseline for comparison, we consider the case where real ictal samples from all other patients and inter-ictal samples from the target patient are available. Therefore, we first build a target and baseline training sets. The target training set consists of 2,000 synthetic ictal samples and 2,000 real inter-ictal samples from the target patient. The baseline training set of 2,000 samples of real seizures randomly selected from all the patients except for the target patient and 2,000 inter-ictal samples from the target patient. In this way, the synthetic ictal samples are strictly the only aspect that differs between the target training set and the baseline training set. Then, for each patient, we construct the test set for the seizure detection task, which contains all the ictal samples of the target patient without overlap and twice as many inter-ictal samples. These test ictal samples have not been used during the training of the GAN (the target patient is left out) to ensure there is no information leakage. We build an unbalanced test set with twice as many inter-ictal as ictal samples in order to better reproduce the real-world setting where the ictal samples are under-represented in the inference phase.

Once the data is split in training and test sets, a feature extraction step is performed on the data following \cite{sopic2018glass} and extract 54 features of power and non-linearity per electrode. Then, both training sets are used to train and evaluate the random forest classifier on the test set. For the sake of robustness, we repeat these experiments 15 times, splitting and shuffling the data each time.

\section{Results}

To asses the performance of the training sets in our experiments, we use the the geometric mean of sensitivity and specificity \citep{fleming1986not}. We report the detailed results of these experiments for each patient in the Appendix \ref{res}. 

\begin{figure}[h]
\centering
\includegraphics[width=.48\linewidth]{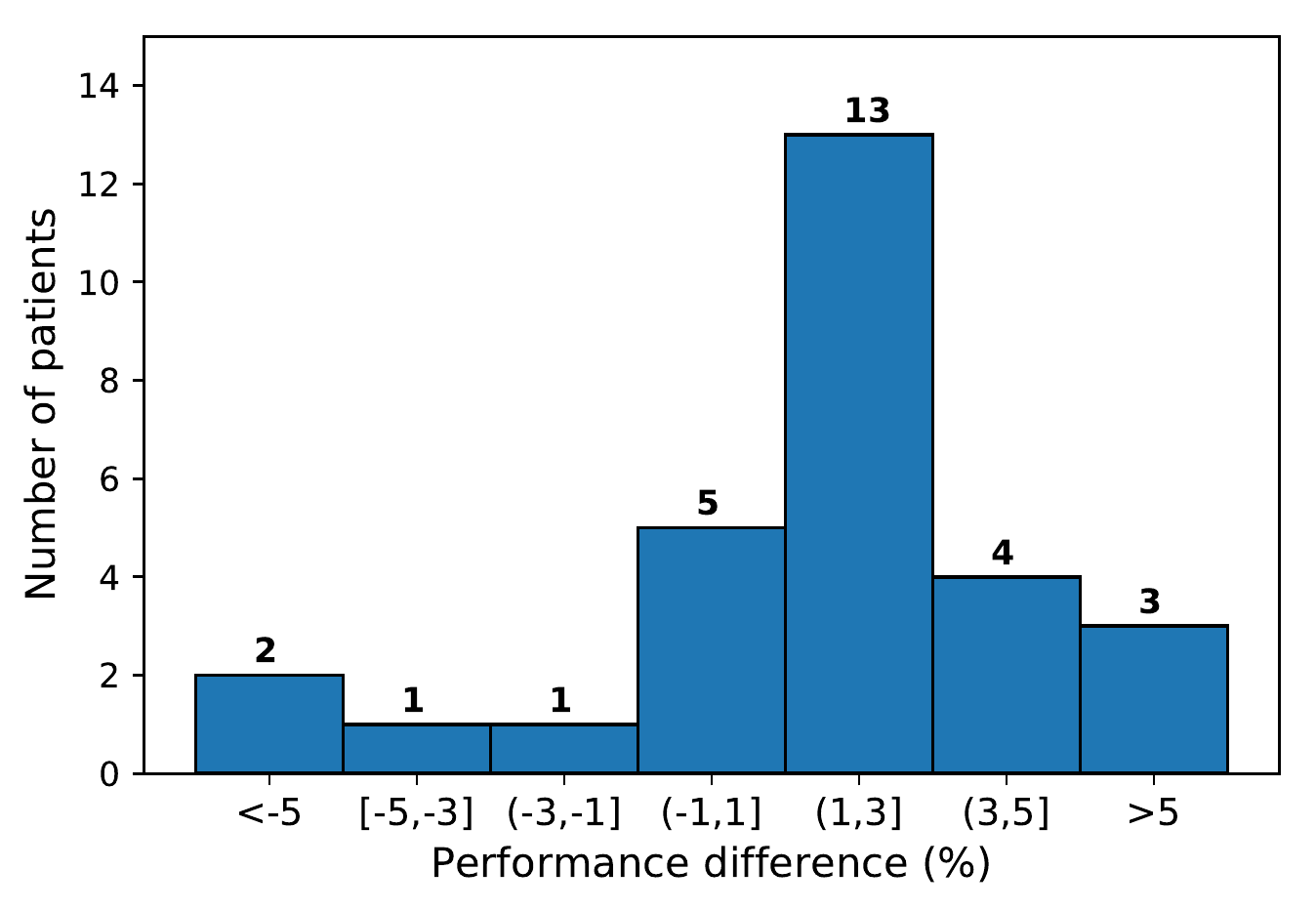}
\caption{Performance difference in the classification task between synthetic and real training sets. The vertical axis represents the number of patients, where larger is better.}
\label{histo}
\end{figure}

Our results show that training with synthetic samples not only does it not degrade the performance, but yields a 1.2\% improvement overall compared to training only with real samples from a generic database. On top of that, as detailed in Figure \ref{histo}, 20 out of 29 patients, i.e., 69\%, improve by more than 1\%, while only for four out of 29 patients the performance decreases by more than 1\%. An explanation for the performance improvement when using synthetic data is that, since our GAN generates ictal samples given inter-ictal samples from the same patient, the model generates synthetic seizures that retain a number of personal features. These results demonstrate the high quality of the synthetic ictal data that we generate as well as its utility for the real-world task of seizure detection.


\section{Conclusion}
In this work, we have presented a GAN model that generates synthetic EEG signals of epileptic seizures. To the best of our knowledge, for the first time in the medical domain, we have generated synthetic data sets that can train detection algorithms achieving results comparable to training with real data. Our results underline that, in the most common scenario in which no recordings of epileptic seizures of a new patient are available, training can be done using exclusively synthetic seizures. Hence, using an existing database, deep generative models can generate data to train a system to monitor a new patient. Our work emphasizes that the application in medicine of deep generative models, such as GANs, can help solving some of the open challenges in the field and help bridging the gap on the adoption of continuous monitoring systems for patients suffering from chronic disorders. Further research into unpaired and conditional deep generative models may improve the quality and performance of synthetic training sets allowing for personalized medicine based on synthetic data.

\bibliographystyle{plainnat.bst}
\bibliography{sample}

\begin{thebibliography}{34}
\providecommand{\natexlab}[1]{#1}
\providecommand{\url}[1]{\texttt{#1}}
\expandafter\ifx\csname urlstyle\endcsname\relax
  \providecommand{\doi}[1]{doi: #1}\else
  \providecommand{\doi}{doi: \begingroup \urlstyle{rm}\Url}\fi

\bibitem[Alotaiby et~al.(2014)Alotaiby, Alshebeili, Alshawi, Ahmad, and
  El-Samie]{alotaiby2014eeg}
Turkey~N Alotaiby, Saleh~A Alshebeili, Tariq Alshawi, Ishtiaq Ahmad, and Fathi
  E~Abd El-Samie.
\newblock Eeg seizure detection and prediction algorithms: a survey.
\newblock \emph{EURASIP Journal on Advances in Signal Processing},
  2014\penalty0 (1):\penalty0 183, 2014.

\bibitem[Aznan et~al.(2019)Aznan, Atapour-Abarghouei, Bonner, Connolly,
  Moubayed, and Breckon]{aznan2019simulating}
Nik Khadijah~Nik Aznan, Amir Atapour-Abarghouei, Stephen Bonner, Jason
  Connolly, Noura~Al Moubayed, and Toby Breckon.
\newblock Simulating brain signals: Creating synthetic eeg data via
  neural-based generative models for improved ssvep classification.
\newblock \emph{arXiv preprint arXiv:1901.07429}, 2019.

\bibitem[Bandt and Pompe(2002)]{bandt2002permutation}
Christoph Bandt and Bernd Pompe.
\newblock Permutation entropy: a natural complexity measure for time series.
\newblock \emph{Physical review letters}, 88\penalty0 (17):\penalty0 174102,
  2002.

\bibitem[Blumenfeld(2012)]{blumenfeld2012impaired}
Hal Blumenfeld.
\newblock Impaired consciousness in epilepsy.
\newblock \emph{The Lancet Neurology}, 11\penalty0 (9):\penalty0 814--826,
  2012.

\bibitem[Che et~al.(2017)Che, Cheng, Zhai, Sun, and Liu]{che2017boosting}
Zhengping Che, Yu~Cheng, Shuangfei Zhai, Zhaonan Sun, and Yan Liu.
\newblock Boosting deep learning risk prediction with generative adversarial
  networks for electronic health records.
\newblock In \emph{2017 IEEE International Conference on Data Mining (ICDM)},
  pages 787--792. IEEE, 2017.

\bibitem[Corley and Huang(2018)]{corley2018deep}
Isaac~A Corley and Yufei Huang.
\newblock Deep eeg super-resolution: Upsampling eeg spatial resolution with
  generative adversarial networks.
\newblock In \emph{2018 IEEE EMBS International Conference on Biomedical \&
  Health Informatics (BHI)}, pages 100--103. IEEE, 2018.

\bibitem[Costa et~al.(2018)Costa, Galdran, Meyer, Niemeijer, Abr{\`a}moff,
  Mendon{\c{c}}a, and Campilho]{costa2018end}
Pedro Costa, Adrian Galdran, Maria~Ines Meyer, Meindert Niemeijer, Michael
  Abr{\`a}moff, Ana~Maria Mendon{\c{c}}a, and Aur{\'e}lio Campilho.
\newblock End-to-end adversarial retinal image synthesis.
\newblock \emph{IEEE transactions on medical imaging}, 37\penalty0
  (3):\penalty0 781--791, 2018.

\bibitem[Debener et~al.(2015)Debener, Emkes, De~Vos, and
  Bleichner]{debener2015unobtrusive}
Stefan Debener, Reiner Emkes, Maarten De~Vos, and Martin Bleichner.
\newblock Unobtrusive ambulatory eeg using a smartphone and flexible printed
  electrodes around the ear.
\newblock \emph{Scientific reports}, 5:\penalty0 16743, 2015.

\bibitem[D{\'\i}az-Uriarte and De~Andres(2006)]{diaz2006gene}
Ram{\'o}n D{\'\i}az-Uriarte and Sara~Alvarez De~Andres.
\newblock Gene selection and classification of microarray data using random
  forest.
\newblock \emph{BMC bioinformatics}, 7\penalty0 (1):\penalty0 3, 2006.

\bibitem[Esteban et~al.(2017)Esteban, Hyland, and R{\"a}tsch]{esteban2017real}
Crist{\'o}bal Esteban, Stephanie~L Hyland, and Gunnar R{\"a}tsch.
\newblock Real-valued (medical) time series generation with recurrent
  conditional gans.
\newblock \emph{arXiv preprint arXiv:1706.02633}, 2017.

\bibitem[Fleming and Wallace(1986)]{fleming1986not}
Philip~J Fleming and John~J Wallace.
\newblock How not to lie with statistics: the correct way to summarize
  benchmark results.
\newblock \emph{Communications of the ACM}, 29\penalty0 (3):\penalty0 218--221,
  1986.

\bibitem[Frid-Adar et~al.(2018)Frid-Adar, Diamant, Klang, Amitai, Goldberger,
  and Greenspan]{frid2018gan}
Maayan Frid-Adar, Idit Diamant, Eyal Klang, Michal Amitai, Jacob Goldberger,
  and Hayit Greenspan.
\newblock Gan-based synthetic medical image augmentation for increased cnn
  performance in liver lesion classification.
\newblock \emph{Neurocomputing}, 321:\penalty0 321--331, 2018.

\bibitem[Goodfellow et~al.(2014)Goodfellow, Pouget-Abadie, Mirza, Xu,
  Warde-Farley, Ozair, Courville, and Bengio]{goodfellow2014}
Ian Goodfellow, Jean Pouget-Abadie, Mehdi Mirza, Bing Xu, David Warde-Farley,
  Sherjil Ozair, Aaron Courville, and Yoshua Bengio.
\newblock Generative adversarial nets.
\newblock In Z.~Ghahramani, M.~Welling, C.~Cortes, N.~D. Lawrence, and K.~Q.
  Weinberger, editors, \emph{Advances in Neural Information Processing Systems
  27}, pages 2672--2680. Curran Associates, Inc., 2014.
\newblock URL
  \url{http://papers.nips.cc/paper/5423-generative-adversarial-nets.pdf}.

\bibitem[Goverdovsky et~al.(2017)Goverdovsky, von Rosenberg, Nakamura, Looney,
  Sharp, Papavassiliou, Morrell, and Mandic]{goverdovsky2017hearables}
Valentin Goverdovsky, Wilhelm von Rosenberg, Takashi Nakamura, David Looney,
  David~J Sharp, Christos Papavassiliou, Mary~J Morrell, and Danilo~P Mandic.
\newblock Hearables: Multimodal physiological in-ear sensing.
\newblock \emph{Scientific reports}, 7\penalty0 (1):\penalty0 6948, 2017.

\bibitem[Hartmann et~al.(2018)Hartmann, Schirrmeister, and
  Ball]{hartmann2018eeg}
Kay~Gregor Hartmann, Robin~Tibor Schirrmeister, and Tonio Ball.
\newblock Eeg-gan: Generative adversarial networks for electroencephalograhic
  (eeg) brain signals.
\newblock \emph{arXiv preprint arXiv:1806.01875}, 2018.

\bibitem[Hirtz et~al.(2007)Hirtz, Thurman, Gwinn-Hardy, Mohamed, Chaudhuri, and
  Zalutsky]{hirtz2007common}
D~Hirtz, DJ~Thurman, K~Gwinn-Hardy, M~Mohamed, AR~Chaudhuri, and R~Zalutsky.
\newblock How common are the “common” neurologic disorders?
\newblock \emph{Neurology}, 68\penalty0 (5):\penalty0 326--337, 2007.

\bibitem[Hoppe et~al.(2015)Hoppe, Feldmann, Blachut, Surges, Elger, and
  Helmstaedter]{hoppe2015novel}
Christian Hoppe, Mieke Feldmann, Barbara Blachut, Rainer Surges, Christian~E
  Elger, and Christoph Helmstaedter.
\newblock Novel techniques for automated seizure registration: patients' wants
  and needs.
\newblock \emph{Epilepsy \& Behavior}, 52:\penalty0 1--7, 2015.

\bibitem[Ihle et~al.(2012)Ihle, Feldwisch-Drentrup, Teixeira, Witon, Schelter,
  Timmer, and Schulze-Bonhage]{ihle2012epilepsiae}
Matthias Ihle, Hinnerk Feldwisch-Drentrup, C{\'e}sar~A Teixeira, Adrien Witon,
  Bj{\"o}rn Schelter, Jens Timmer, and Andreas Schulze-Bonhage.
\newblock Epilepsiae--a european epilepsy database.
\newblock \emph{Computer methods and programs in biomedicine}, 106\penalty0
  (3):\penalty0 127--138, 2012.

\bibitem[Kingma and Ba(2014)]{kingma2014adam}
Diederik~P Kingma and Jimmy Ba.
\newblock Adam: A method for stochastic optimization.
\newblock \emph{arXiv preprint arXiv:1412.6980}, 2014.

\bibitem[Klem et~al.(1999)Klem, L{\"u}ders, Jasper, Elger, et~al.]{klem1999ten}
George~H Klem, Hans~Otto L{\"u}ders, HH~Jasper, C~Elger, et~al.
\newblock The ten-twenty electrode system of the international federation.
\newblock \emph{Electroencephalogr Clin Neurophysiol}, 52\penalty0
  (3):\penalty0 3--6, 1999.

\bibitem[Lipton et~al.(2016)Lipton, Kale, and Wetzel]{lipton2016modeling}
Zachary~C Lipton, David~C Kale, and Randall Wetzel.
\newblock Modeling missing data in clinical time series with rnns.
\newblock \emph{Machine Learning for Healthcare}, 2016.

\bibitem[Maas et~al.(2013)Maas, Hannun, and Ng]{maas2013rectifier}
Andrew~L Maas, Awni~Y Hannun, and Andrew~Y Ng.
\newblock Rectifier nonlinearities improve neural network acoustic models.
\newblock In \emph{Proc. icml}, volume~30, page~3, 2013.

\bibitem[Mao et~al.(2017)Mao, Li, Xie, Lau, Wang, and
  Paul~Smolley]{Mao_2017_ICCV}
Xudong Mao, Qing Li, Haoran Xie, Raymond~Y.K. Lau, Zhen Wang, and Stephen
  Paul~Smolley.
\newblock Least squares generative adversarial networks.
\newblock In \emph{The IEEE International Conference on Computer Vision
  (ICCV)}, Oct 2017.

\bibitem[Mirza and Osindero(2014)]{mirza2014conditional}
Mehdi Mirza and Simon Osindero.
\newblock Conditional generative adversarial nets.
\newblock \emph{arXiv preprint arXiv:1411.1784}, 2014.

\bibitem[Miyato et~al.(2018)Miyato, Kataoka, Koyama, and
  Yoshida]{Miyato18spectral}
Takeru Miyato, Toshiki Kataoka, Masanori Koyama, and Yuichi Yoshida.
\newblock Spectral normalization for generative adversarial networks.
\newblock In \emph{6th International Conference on Learning Representations,
  {ICLR} 2018, Vancouver, BC, Canada, April 30 - May 3, 2018, Conference Track
  Proceedings}, 2018.

\bibitem[Osorio et~al.(2016)Osorio, Zaveri, Frei, and
  Arthurs]{osorio2016epilepsy}
I.~Osorio, H.P. Zaveri, M.G. Frei, and S.~Arthurs.
\newblock \emph{Epilepsy: The Intersection of Neurosciences, Biology,
  Mathematics, Engineering, and Physics}.
\newblock CRC Press, 2016.
\newblock ISBN 9781439838860.
\newblock URL \url{https://books.google.ch/books?id=O97hKvyyYgsC}.

\bibitem[Pascual et~al.(2017)Pascual, Bonafonte, and
  Serr{\`a}]{pascual2017segan}
Santiago Pascual, Antonio Bonafonte, and Joan Serr{\`a}.
\newblock Segan: Speech enhancement generative adversarial network.
\newblock In \emph{INTERSPEECH}, 2017.

\bibitem[Price and Cohen(2019)]{price2019privacy}
W~Nicholson Price and I~Glenn Cohen.
\newblock Privacy in the age of medical big data.
\newblock \emph{Nature medicine}, 25\penalty0 (1):\penalty0 37, 2019.

\bibitem[Richman and Moorman(2000)]{richman2000physiological}
Joshua~S Richman and J~Randall Moorman.
\newblock Physiological time-series analysis using approximate entropy and
  sample entropy.
\newblock \emph{American Journal of Physiology-Heart and Circulatory
  Physiology}, 278\penalty0 (6):\penalty0 H2039--H2049, 2000.

\bibitem[Ronneberger et~al.(2015)Ronneberger, Fischer, and
  Brox]{ronneberger2015u}
Olaf Ronneberger, Philipp Fischer, and Thomas Brox.
\newblock U-net: Convolutional networks for biomedical image segmentation.
\newblock In \emph{International Conference on Medical image computing and
  computer-assisted intervention}, pages 234--241. Springer, 2015.

\bibitem[Salimans et~al.(2016)Salimans, Goodfellow, Zaremba, Cheung, Radford,
  and Chen]{salimans2016improved}
Tim Salimans, Ian Goodfellow, Wojciech Zaremba, Vicki Cheung, Alec Radford, and
  Xi~Chen.
\newblock Improved techniques for training gans.
\newblock In \emph{Advances in neural information processing systems}, pages
  2234--2242, 2016.

\bibitem[Shin et~al.(2018)Shin, Tenenholtz, Rogers, Schwarz, Senjem, Gunter,
  Andriole, and Michalski]{shin2018medical}
Hoo-Chang Shin, Neil~A Tenenholtz, Jameson~K Rogers, Christopher~G Schwarz,
  Matthew~L Senjem, Jeffrey~L Gunter, Katherine~P Andriole, and Mark Michalski.
\newblock Medical image synthesis for data augmentation and anonymization using
  generative adversarial networks.
\newblock In \emph{International Workshop on Simulation and Synthesis in
  Medical Imaging}, pages 1--11. Springer, 2018.

\bibitem[Sopic et~al.(2018)Sopic, Aminifar, and Atienza]{sopic2018glass}
Dionisije Sopic, Amir Aminifar, and David Atienza.
\newblock {e-Glass: A wearable system for real-time detection of epileptic
  seizures}.
\newblock In \emph{2018 IEEE International Symposium on Circuits and Systems
  (ISCAS)}, pages 1--5. IEEE, 2018.

\bibitem[Sperling et~al.(1999)Sperling, Feldman, Kinman, Liporace, and
  O'Connor]{sperling1999seizure}
Michael~R Sperling, Harold Feldman, Judith Kinman, Joyce~D Liporace, and
  Michael~J O'Connor.
\newblock Seizure control and mortality in epilepsy.
\newblock \emph{Annals of neurology}, 46\penalty0 (1):\penalty0 45--50, 1999.

\end{thebibliography}

\clearpage
\appendix

\section{Results}\label{res}

Table~\ref{tab} contains the per patient results of our experiments. Patient 22 performs extremely poorly for both the baseline and the synthetic training sets and, therefore, it is not a relevant indicator of the classification quality. Consequently, it has been removed from the calculation of the total difference in performance. This total difference is calculated as the difference between the geometric mean of all patients in the synthetic case and the geometric mean of all patients in the baseline case. These results pass the Wilcoxon statistical significance test with a \textit{p}-value of 0.0098 when patient 22 is already excluded, which indicates that the difference between the results obtained for the baseline and synthetic training sets is statistically significant.

\begin{table}[h]
\centering
\begin{tabular}{|c|c|c|c|}
\hline
Patient ID & Baseline (\%) & Synthetic (\%) & Difference (\%)\\
\hline
1 & 73.49 & 80.06 & \textbf{+6.57}\\ %
\hline
2 & 79.36 & 70.30 & -9.06\\
\hline
3 & 77.59 & 82.84 & \textbf{+5.25}\\
\hline
4 & 76.34 & 78.33 & \textbf{+1.99}\\
\hline
5 & 64.86 & 68.19 & \textbf{+3.33}\\ %
\hline
6 & 74.10 & 74.74 & +0.64\\
\hline
7 & 68.11 & 68.59 & +0.48\\
\hline
8 & 81.41 & 86.14 & \textbf{+4.73}\\
\hline
9 & 76.74 & 80.67 & \textbf{+3.93}\\
\hline
10 & 66.84 & 65.87 & \textbf{-0.97}\\
\hline
11 & 81.03 & 83.66 & \textbf{+2.63}\\
\hline
12 & 63.00	& 66.56 & \textbf{+3.56}\\
\hline
13 & 77.20	& 78.54 & \textbf{+1.34}\\
\hline
14 & 74.32	& 76.51 & \textbf{+2.19}\\
\hline
15 & 74.25	& 74.07 & -0.18\\
\hline
16 & 78.11	& 80.64 & \textbf{+2.53}\\
\hline
17 & 65.27	& 67.84 & \textbf{+2.57}\\
\hline
18 & 66.20	& 71.62 & \textbf{+5.42}\\
\hline
19 & 76.95	& 78.13 & \textbf{+1.18}\\
\hline
20 & 73.42	& 68.67 & -4.75\\
\hline
21 & 79.18	& 71.61 & -7.57\\
\hline
\st{\textit{22}} & \st{\textit{26.88}} & \st{\textit{12.62}} & \st{\textit{-14.26}}\\
\hline
23 & 77.05	& 78.62 & \textbf{+1.57}\\
\hline
24 & 78.28	& 77.87 & -0.41\\
\hline
25 & 77.02	& 75.65 & -1.37\\
\hline
26 & 74.36	& 76.15 & \textbf{+1.79}\\
\hline
27 & 76.00	& 78.00 & \textbf{+2.0}\\
\hline
28 & 81.97	& 83.07 & \textbf{+1.10}\\
\hline
29 & 75.73	& 78.41 & \textbf{+2.68}\\
\hline
30 & 79.80	& 82.29 & \textbf{+2.49}\\
\hline
TOTAL & 74.57 & 75.78 & \textbf{+1.21}\\
\hline
\end{tabular}
\caption{\label{tab} Geometric mean of sensitivity and specificity per patient of our evaluation.}
\end{table}

Regarding the patients for whom the performance degrades most significantly, i.e., Patients 2 and 21, their seizures are dominated by repetitive spiking. This pattern is relatively rare and is not well represented in the dataset (only 10.5\% of the seizures). Therefore, our GAN model does not capture this behavior with as much precision as it does capture other patterns such as theta or delta rhythms. In fact, Patient 22 also suffers from seizures with repetitive spiking and our experiments show that even the state-of-the-art techniques fail in detection of such seizures. Finally, Patient 20 has only 4 seizures in this dataset, which is the lowest number of seizures in the entire dataset and hinders robust evaluation of our model.

\section{GAN architectural details}
The architecture of our GAN is modeled after the SEGAN from \citep{pascual2017segan}. Our generator is a U-net \citep{ronneberger2015u} convolutional autoencoder with weighted skip connections in all the layers except for the input and the latent code layers. The weights of the skip connections are learnt during training. The input to the generator are samples of length 2048 points (4 seconds of signal from 2 electrodes recorded at a frequency of 256 Hz). The encoder consists of eight blocks that alternate a convolutional layer with a max pooling layer with 2x2 filters and stride 2. The feature maps extracted at each block of the encoder yield the following shapes: 2048x1, 1024x64, 512x64, 256x128, 128x128, 64x256, 32x256, 16x512, 8x1024; where 2048x1 is the shape of the input and 8x1024 that of the latent encoding. Gaussian noise with mean 0 and standard deviation 1 and shape 8x1024 is concatenated to the latent code in order to introduce stochasticity into the model. The decoder is symmetric to the encoder, but it uses deconvolutions and dilations. Thus, the shapes of the feature maps at the decoder are 16x1024, 16x512, 32x256, 64x256, 128x128, 256x128, 512x64, 1024x64, 2048x1; where 2048x1 is the final output of the generator. The skip connections multiply the feature maps output at each layer of the encoder with the learned weights and adds it to the feature map of the same shape at the decoder. The activation used is the leakyReLu function \citep{maas2013rectifier}, except for the last block of the decoder, where we use the hyperbolic tangent function. All convolutions and deconvolutions are unbiased and spectral normalization \citep{Miyato18spectral} is applied before each block. 

The architecture of the discriminator has the same shape as the encoder of the generator, but it includes an additional fully connected layer at the output. In this way, the discriminator outputs a single value between 1 and 0, where 1 represents the real class and 0 the synthetic class. On top of that, in the discriminator we apply virtual batch normalization \citep{salimans2016improved}, as well as spectral normalization.

To train the model we use the Adam \citep{kingma2014adam} optimizer with 0 and 0.9 for the values of $\beta_1$ and $\beta_2$, respectively, and learning rates 0.0001 for the generator and 0.0004 for the discriminator. The size of the minibatches of data employed during training is 100 samples. All the hyperparameters employed are summarized in Table \ref{param}.

\begin{table}[ht]
\centering
\begin{tabular}{|c|c|}
\hline
Parameter & Value\\
\hline
$\beta_1$ & 0\\ %
\hline
$\beta_2$ & 0.9\\
\hline
G learning rate & 0.0001\\
\hline
D learning rate & 0.0004\\
\hline
$\lambda$ & 100\\ 
\hline
Mini-batch size & 100\\ 
\hline
\end{tabular}
\caption{\label{param} Model hyperparameters.}
\end{table}

\section{Evaluation details}. 

In the feature extraction stage we follow \cite{sopic2018glass} and extract 54 features of power and non-linearity per electrode, i.e., a total of 108 features. To calculate the non-linear features, the signal is decomposed using the discrete wavelet transform down to level seven. The nonlinear features extracted are: sixth and seventh level sample entropy \citep{richman2000physiological} for $k = 0.2$ and $k = 0.35$; third, fourth, fifth, sixth and seventh level permutation entropy \citep{bandt2002permutation} for $n=3$, $n=5$ and $n=7$; third, fourth, fifth, sixth and seventh level, as well as raw signal, Shannon, Renyi and Tsallis entropies. The power features are: total power, total and relative band power in the bands delta [0.5,4] Hz, theta [4,8] Hz, alpha [8,12] Hz, beta [13,30] Hz, gamma [30,45] Hz as well as in the bands [0,0.1] Hz, [0.1,0.5] Hz, [12,13] Hz.

\end{document}